Original Ultra-hard Orthorhombic Carbon allotropes C$_8$ with lonsdaleite-type **lon** topology: Crystallographic and DFT investigations.

Samir F. Matar

Lebanese German University (LGU), Sahel Alma, P. O. Box 206 Jounieh, Lebanon.

https://orcid.org/0000-0001-5419-358X

**Abstract.**

The **lon** topology inherent to lonsdaleite C$_4$ (hexagonal diamond) is shown to characterize three original carbon allotropes C$_8$ in the orthorhombic system with irregular orientations of the *C4* tetrahedra. The octacarbon stoichiometries were devised from crystal structure engineering and identified close to lonsdaleite from density functional theory (DFT) based calculations of ground state structures and energy derived physical properties. Characterized with high densities, the three allotropes are identified with ultra-hard with hardness magnitudes close and superior to lonsdaleite. Dynamically, all three allotropes were found stable with positive frequencies revealed from their phonons band structures with specific heat C$_V$ = $f$(T) calculated curves close to diamond's experimental values from literature.

**Introduction**

The field of carbon research occupies a particular position among scientists of the solid-state, especially with the search of allotropes related to diamond's exceptional physical properties. Also, original artificial allotropes with similar mechanical and thermal properties to diamond are regularly proposed based on crystal structure engineering [1] or using structure prediction codes as USPEX [2]. A library of such original and mainly artificial carbon edifices was established for storing the devised structures, namely the SACADA database [3, 4]. The classification is based on topology analysis using TopCryst program [5]. For instance, diamond is identified with **dia** label while its rare hexagonal form lonsdaleite first announced by Bundy and Kasper back in 1967 [6] is labeled **lon**. Questioning its very existence, Németh et al. declared in 2014 that lonsdaleite was actually "*a faulted and twinned cubic diamond and does not exist as a discrete material*" [7]. The **lon** topology occurs in SACADA four times: No. 37 (lonsdaleite itself), 38, 128, and **lon-a** for No. 1189. As lonsdaleite, all **lon** carbon allotropes crystallize in the hexagonal system with $P6_3/mmc$ No. 194 and $P\text{-}6m2$ No. 186 space groups.

Regarding the density criterion, both diamond and lonsdaleite possess high densities of $\rho \sim 3.52$ g/cm$^3$ arising from the perfectly covalent character of the C-C short connections (1.54 Å) within C(sp$^3$)-like tetrahedral carbon resulting in a high (Vickers) hardness $H_V \geq 95$ GPa. However, based on first principles calculations lonsdaleite was announced stiffer and harder than diamond [8]. Such quantum mechanics computations usually carried out in the framework of the widely adopted Density Functional Theory (DFT) framework [9,10] are a powerful predictive tool of ground state structures and energy derived properties as the physical ones (elastic properties, phonons, electronic band structures, …). For instance, an artificial carbon allotrope with **qtz** topology (i.e., quartz-based) was announced as denser than diamond, with $\rho = 3.67$ g/cm$^3$ and subsequent higher Vickers hardness $H_V \sim 100$ GPa, (cf. [11] and therein cited works). Such properties arise from the structural local modifications as the presence of strongly distorted *C4* tetrahedra and the manner they are connected; mainly corner sharing (cf. Fig. 1a relevant to diamond). Nevertheless, such systems are considered as metastable versus diamond despite their computed mechanical, dynamic, thermodynamic stabilities. Considering lonsdaleite shown in Fig. 1b, it is characterized by four atoms at 4*f* position of the high symmetry space group $P6_3/mmc$ No. 194 (Table 1). In preliminary trials, it was noted that resolving the structure in orthorhombic symmetry through crystallography engineering using the lonsdaleite atomic coordinates, a trend of transformation into 8 atoms $C_8$ systems with 3D tetrahedral structure was noticed. This characterizes the presently proposed original orthorhombic $C_8$ structures in three different orthorhombic space groups. The crystallographic findings were necessarily backed with DFT investigations of the ground state structures and the energy dependent physical properties. The orthorhombic $C_8$ allotropes were identified with **lon** topology known, as discussed above, to characterize hexagonal carbon allotropes. Our interest herein is not to question the existence of lonsdaleite as such but to show that **lon** topology cal also be found in another symmetry than hexagonal systems with stackings of *C4* tetrahedra, different from lonsdaleite.

After this Introduction, the paper is organized as follows: the Theoretical framework and the Computational methodology are given in Section 1; the Crystal Structure characteristics are presented in Section 2; the Mechanical properties from the elastic constants are addressed in Section 3; the Dynamic properties from the Phonons are detailed in Section 4 and the pertaining Temperature dependence of the heat capacity in comparison with Diamond experimental data is given in Section 5. Section 6 presents the Electronic band structures. The paper is ended with a Conclusion.

### 1- Theoretical framework and Computational methodology

To determine the ground state structures corresponding to energy minima and to derive the mechanical, dynamic properties and the electronic structures, quantum mechanics computations were carried out based on the widely adopted framework of the density functional theory DFT [9,10]. Based on the DFT, the calculations were performed using the Vienna Ab initio Simulation Package (VASP) code [12, 13] and the Projector Augmented Wave (PAW) method [13, 14] for the atomic potentials. DFT exchange-correlation (XC) effects were considered using the generalized gradient approximation (GGA) [15]. Relaxation of the atoms onto the ground state structures was performed with the conjugate gradient algorithm according to Press *et al*. [16]. The Bloechl tetrahedron method [17] with corrections according to the scheme of Methfessel and Paxton [18] was used for geometry optimization and energy calculations. Brillouin-zone (BZ) integrals were approximated by a special **k**-point sampling according to Monkhorst and Pack [19]. Structural parameters were optimized until atomic forces were below 0.02 eV/Å and all stress components < 0.003 eV/Å$^3$. The calculations were converged at an energy cutoff of 400 eV for the plane-wave basis set in terms of the automatic high precision **k**-point integration in the reciprocal space to obtain a final convergence and relaxation to zero strains for the original stoichiometries presented in this work. In the post-processing of the ground state electronic structures, the charge density projections were operated on the lattice sites.

The mechanical stabilities were obtained from the calculations of the elastic constants Cij subsequently treated with ELATE online program [20]. The outcome provides the bulk (B) and shear (G) modules along different averaging methods; the Voigt method [21] was adopted herein for $B_V$ and $G_V$. The methods of microscopic theory of hardness by Tian et al. [22] and Chen et al. [23] were used to estimate the Vickers hardness ($H_V$) from the bulk and shear modules $B_V$ and $G_V$ (*vide infra*).

The dynamic stabilities were confirmed by the calculation of the phonons band structures all presenting positive phonon frequencies. The corresponding phonon band structures were obtained from a high resolution of the orthorhombic Brillouin zone according to Togo *et al*. [24].

Experimental specific heat $C_V$ data of diamond needed to assess the calculated results of the three 3D allotropes were obtained from Victor works [25]. The electronic band structures were obtained using the all-electron DFT-based ASW method [26] and the GGA XC functional [15]. The VESTA (Visualization for Electronic and Structural Analysis) program [27] was used to visualize the crystal structures.

### 2- Crystal structures

Despite the close structural properties of both diamond (cubic) and lonsdaleite (hexagonal) being built-up with *C4* tetrahedra in a three-dimensional network, they exhibit differences regarding the stacks of tetrahedra: they are oriented in the same direction in diamond and alternately oriented in lonsdaleite. These differentiating aspects are illustrated in Figures 1a and 1b. The original orthorhombic $C_8$ (Figs. 1c – 1d) showing these features are identified in orthorhombic space groups *C*222$_1$ No. 20, *Pmc*2$_1$ No. 26, and *Ama*2 No. 40. From Table 1, lonsdaleite calculated *a* and *c* lattice parameters show differences with respect to Bundy and Kasper 1967 paper values with a smaller *c* lattice constant [6] but they remain within accepted range. The crystal structure characteristics of the $C_8$ allotropes are close to lonsdaleite regarding the interatomic distances, the *C4* tetrahedron angle of 109°, the atom average volume close to 5.67 Å$^3$, and the density ρ = 3.52 g/cm$^3$. Note that $C_8$ $Pmc2_1$ No. 26 shows two sites for the atomic positions that are related to each other with a translation ½ ½ ±x, however a search for a higher symmetry orthorhombic space group was not successful. Regarding the energy, the atom averaged cohesive magnitudes are the same and equal to $E_{coh}$/at.(diamond) = -2.47 eV. Finally, the *Topcryst* program analysis showed that the topological mapping is **lon**.

From the crystal structure assessments, physical properties close to lonsdaleite can be expected. The following sections of the paper are devoted to further analyze the mechanical, dynamic thermodynamic, and electronic structures properties of the three $C_8$ allotropes carried out in comparison to lonsdaleite.

### 3- Mechanical properties from the elastic constants

The investigation of the mechanical properties was based on the calculations of the elastic properties determined by performing finite distortions of the lattice and deriving the elastic constants from the strain-stress relationship. The calculated sets of elastic constants $C_{ij}$ (i and j indicate directions) are given in Table 2. All $C_{ij}$ values are positive signaling stability of the three allotropes as well as lonsdaleite. The products obey the stability rules regarding the orthorhombic system:

$C_{ii}$ (i =1, 4, 5, 6) > 0.

$C_{11}C_{22} - C_{12}^2 > 0$; $C_{11}C_{22}C_{33} + 2C_{12}C_{13}C_{23} - C_{11}C_{23}^2 - C_{22}C_{13}^2 - C_{33}C_{12}^2 > 0$.

Using ELATE program introduced above [20], the bulk and the shear modules obtained by averaging the elastic constants using Voigt's [21] method are given in Table 3, namely $B_V$ and $G_V$. Table 3 displays the mechanical properties. Large similar magnitudes of the bulk modulus $B_V$ ~440 GPa are observed for lonsdaleite $C_4$ and the other $C_8$ allotropes. From literature Wang [28] announced a magnitude of 434 GPa using GGA DFT XC functional as in the present work. Similarly, the shear moduli are systematically larger. Note that $B_V$ and $G_V$ are within range of diamond's: $B_V$ =444 GPa and $G_V$= 534 GPa [29]. The corresponding Pugh ratios $G_V/B_V$ [30] relevant to "ductile-to-brittle" criteria were then calculated. For $G_V/B_V$ < 1 a trend to ductile behavior is expected whereas for $G_V/B_V$ > 1 a brittle behavior is predicted. All Pugh ratios are found larger than unity letting deduct significant brittleness within range of the value of diamond: $G_V/B_V$ = 1.20. Note that the largest ratio is obtained for $C_8$ $Pmc2_1$. Such magnitudes of Pugh ratios are translated into the Vickers hardness Hv calculated along with two models of microscopic theory of hardness:

$H_V^1 = 0.92(G_V/B_V)^{1.137} G_V^{0.708}$ (Tian et al.) [22]

$H_V^2 = 2(G_V^3/B_V^2)^{0.585} - 3$ (Chen et al.) [23]

Both equations show similar magnitudes of Vickers hardness with an exceptionally high value for $C_8$ $Pmc2_1$ as it can be expected from the large Pugh ratio, followed by $C_8$ $C222_1$. Consequently, it can be concluded that the three **lon** orthorhombic allotropes follow closely lonsdaleite with slightly better mechanical properties of ultra-hardness.

### 4- Dynamic and thermodynamic properties

To verify the dynamic stability of the $C_8$ orthorhombic carbon allotropes, an analysis of their phonon properties was performed concomitantly with lonsdaleite. The phonon band structures were obtained from a high resolution of the hexagonal Brillouin zone BZ for lonsdaleite and the orthorhombic BZ for the three $C_8$ allotropes, according to Togo *et al*. method [24]. Figure 2 shows them in four panels. The bands (red lines) develop along the main directions of the orthorhombic Brillouin zone (horizontal *x*-axis), separated by vertical lines for better visualization, while the vertical direction (*y*-axis) represents the frequencies ω, given in terahertz (THz).

For each crystal system the phonons band structures include 3N bands (N number of atoms) describing three acoustic modes starting from zero energy (ω = 0) at the Γ point, i.e. BZ center, and reaching up to a few terahertz THz, and 3N-3 optical modes at higher energies. The low-frequency acoustic modes are associated with the rigid translation modes of the crystal lattice, i.e., two transverse and one longitudinal. The phonon frequencies are found all positive, indicating that the allotropes are dynamically stable, alike lonsdaleite. However, a slight trend to negative acoustic frequencies at BZ center is observed for $C_8$ $C222_1$ (Fig. 2b) that were not removed with higher resolution BZ integration. This does not deem the allotrope as unstable and shows that the novel allotropes whilst being close to lonsdaleite, are different from it.

The highest bands are observed around 40 THz for lonsdaleite $C_4$ as well as for the orthorhombic $C_8$ allotropes. Note that ω = 40 THz magnitude is close to the value observed for diamond by Raman spectroscopy [31], letting expect relationship with diamond for all systems. Such an assumption needs to be supported by further analysis of the thermal properties of the three allotropes in comparison to diamond.

### 5- Temperature dependence of the heat capacity

The thermodynamic properties were calculated from the phonon frequencies using the statistical thermodynamic approach [32] on a high-precision sampling mesh in the orthorhombic BZ. The temperature dependencies of the entropy S and the heat capacity $C_V$ are presented in four panels in Figure 3. We also report the available experimental $C_V$ discrete data for diamond [25]. The entropy (black curve) increases with temperature as expected from increased disorder with T. The calculated $C_V = f(T)$ curves increase almost linearly up to ~400K then there is a curving down for higher temperatures. Considering experimental data, the curves follow the discrete points of diamond with a better fit observed for $C_8$ allotropes than for lonsdaleite. In view of the closest resemblance with diamond from the point of view of the most relevant characteristic, namely the extreme hardness shown by $C_8$ allotropes. Such resemblance should be illustrated by the electronic band structure.

### 6- Electronic band structures

Using the crystal parameters in Table 1, the electronic band structures were obtained for lonsdaleite and the three carbon allotropes using the all-electrons DFT-based augmented spherical wave method (ASW) [26] and GGA XC approximation [15]. The band structures are displayed in Figure 4.

The bands develop along the main directions of the primitive orthorhombic Brillouin zone. Along the vertical direction all three panels exhibit an energy gap signaling a semi-conducting to insulating behavior. The zero energy is then considered $E_V$, i.e. at the top of the valence band (V) separated from the higher energy conduction band (C). In Fig. 4a representing archetype **lon** $C_4$ the band gap is large with ~5 eV and indirect as it occurs between $\Gamma_V$ and $K_C$ of the hexagonal BZ ($\Gamma$ designates BZ center). The $C_8$ allotropes in Fig. 4b ($C222_1$) and Fig. 4c ($Pmc2_1$) also show large band gaps of similar magnitude with an indirect character. Lastly, in $C_8$ *Ama*2 (Fig. 4d) the large gap is of direct character occurring between $\Gamma_V$ and $\Gamma_C$, of the orthorhombic BZ. Then the three C8 allotrope show close albeit different electronic band structures versus lonsdaleite.

**Conclusion**

Based on quantum mechanics calculations of ground state crystal structures and pertaining physical properties, we presented three novel orthorhombic carbon allotropes with $C_8$ stoichiometry, found to be structurally related to lonsdaleite regarding the stacking of the *C4* tetrahedra. Their subsequent identification in the **lon** topology usually found for hexagonal carbon systems is an original feature. Specifically, the structures were found with high densification leading to ultra-hard behaviors alike lonsdaleite and diamond. Dynamically, the allotropes were found stable with positive frequencies revealed from their phonons band structures with highest frequencies of 40 THz characteristic of diamond Raman spectroscopy from literature. The thermodynamic properties showed specific heat $C_V$ curves in agreement with diamond's literature experimental values, with better agreement than lonsdaleite. It is proposed from this work, a holistic interrelationship: "crystal structure ↔ mechanical ↔ dynamic ↔ electronic properties" for carbon hard materials (*material = structure + properties*).

**Author Contribution:** Conceptualization, methodology, investigation, formal analysis, visualization, writing the paper. No A.I. nor other web resources were used.

**Funding:** This research received no external funding.

**Data Availability Statement:** The data presented in this study are available upon reasonable request due to LGU university restrictions, i.e., privacy, etc.

**Acknowledgments:** Computational facilities from the Lebanese German University are gratefully acknowledged.

**Conflicts of Interest:** The author declares no conflict of interest.

**Tables**

Table 1. Crystal structural properties of carbon allotropes with **lon** topology.

| Space group | $P6_3/mmc$ No.194 $C_4$ lonsdaleite | $C222_1$ No.20 $C_8$ | $Pmc2_1$ No.26 $C_8$ | $Ama2$ No.40 $C_8$ |
|---|---|---|---|---|
| a, Å | 2.505 (2.51 [6]) | 4.3384 | 4.1691 | 4.1690 |
| b, Å | - | 2.5055 | 2.5050 | 4.3387 |
| c, Å | 4.168 (4.12 [6]) | 4.1690 | 4.3144 | 2.5055 |
| Shortest dist. Å | 1.54 | 1.54 | 1.54 | 1.54 |
| Angle ∠C-C-C ° | 109.47 | 109.1 | 109.9 | 109.06 |
| Volume, Å$^3$ | 22.66 | 45.32 | 45.31 | 45.31 |
| V/at., Å$^3$ | 5.67 | 5.67 | 5.66 | 5.66 |
| Density ρ (g/cm$^3$) | 3.5 | 3.52 | 3.52 | 3.52 |
| Atomic positions | C1 (4$f$) 1/3, 2/3, 0.0627 | C1 (8$c$) 0.1667, ½, 0.8128 | C1 (4$c$) 0.6872, ¾ 0.9139<br>C2 (4$c$) 0.1872, ¼ 0.0806 | C1 (8$c$) 0.5628, 0.8333, 0.4 |
| $E_{total}$, eV | -36.28 | -72.6 | -72.55 | -72.55 |
| $E_{tot.}$/at., eV | -9.07 | -9.08 | -9.06 | -9.06 |
| $E_{coh}$/at., eV | -2.48 | -2.48 | -2.47 | -2.47 |

Table 2. Orthorhombic $C_8$ and Lonsdaleite $C_4$. Elastic Constants $C_{ij}$ in GPa.

| $C_{ij}$ | $C_{11}$ | $C_{22}$ | $C_{12}$ | $C_{13}$ | $C_{33}$ | $C_{44}$ | $C_{55}$ | $C_{66}$ |
|---|---|---|---|---|---|---|---|---|
| *C*$222_1$ | 1211 | 1206 | 104 | 14 | 1327 | 551 | 464 | 464 |
| *Pmc*$2_1$ | 1313 | 1208 | 3 | 4 | 1204 | 464 | 550 | 466 |
| *Ama*2 | 1329 | 1206 | 15 | 15 | 1205 | 464 | 550 | 465 |
| $C_4$*(pw)* | 1192 | - | 123 | 14 | 1327 | 553 | - | 465 |

Table 3. Orthorhombic $C_8$ and Lonsdaleite $C_4$. Voigt-average properties of bulk $B_V$ and shear $G_V$ modules and hardness magnitudes along to models (GPa units).

| | $B_V$ | $G_V$ | $G_V/B_V$ | $H_V^1$ | $H_V^2$ |
|---|---|---|---|---|---|
| *C*$222_1$ | 445 | 537 | 1.21 | 98 | 97 |
| *Pmc*$2_1$ | 437 | 537 | 1.23 | 99 | 100 |
| *Ama*2 | 445 | 530 | 1.19 | 95 | 95 |
| $C_4$ *(pw)* | 444 | 534 | 1.17 | 97 | 96 |

Figures

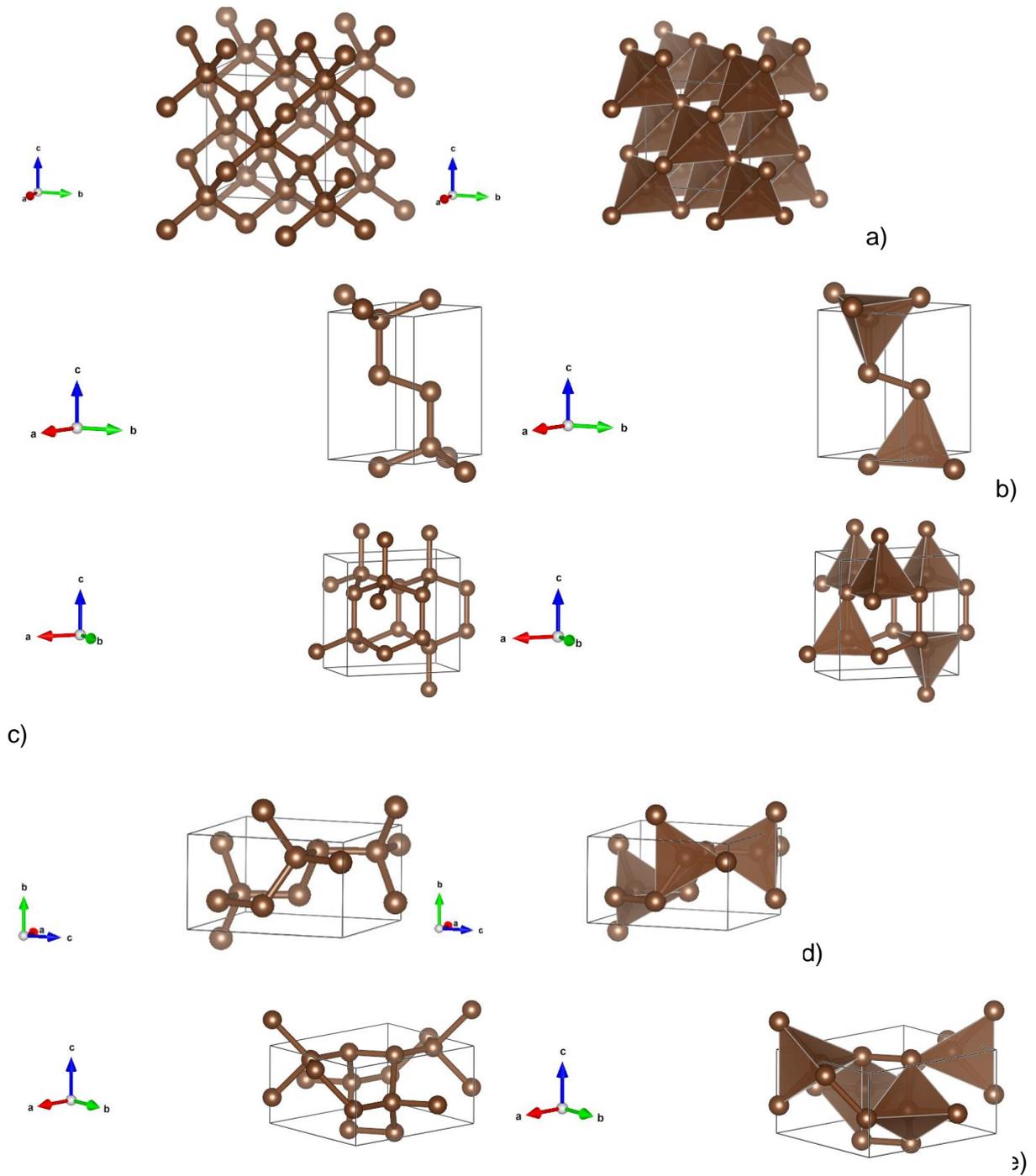

Figure 1. Crystal structures in ball and stick and polyhedral representations of a) Diamond $C_8$, b) Lonsdaleite $C_4$, and the orthorhombic $C_8$ allotropes with **lon** topology designated with their respective space groups: c) $C222_1$, d) $Pmc2_1$, and e) $Ama2$.

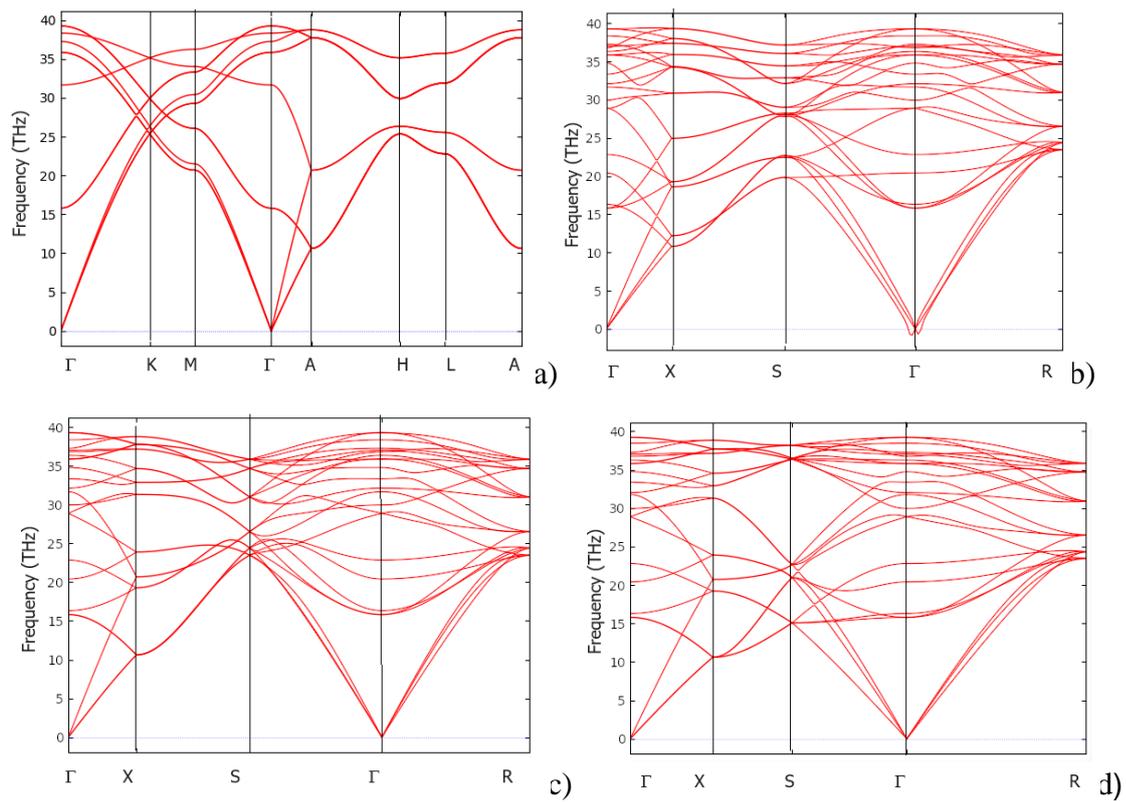

Figure 2. Phonon band structures along the major lines of the orthorhombic Brillouin zone. a) Lonsdaleite $C_4$, and orthorhombic $C_8$: b) $C222_1$, c) $Pmc2_1$, and d) $Ama2$.

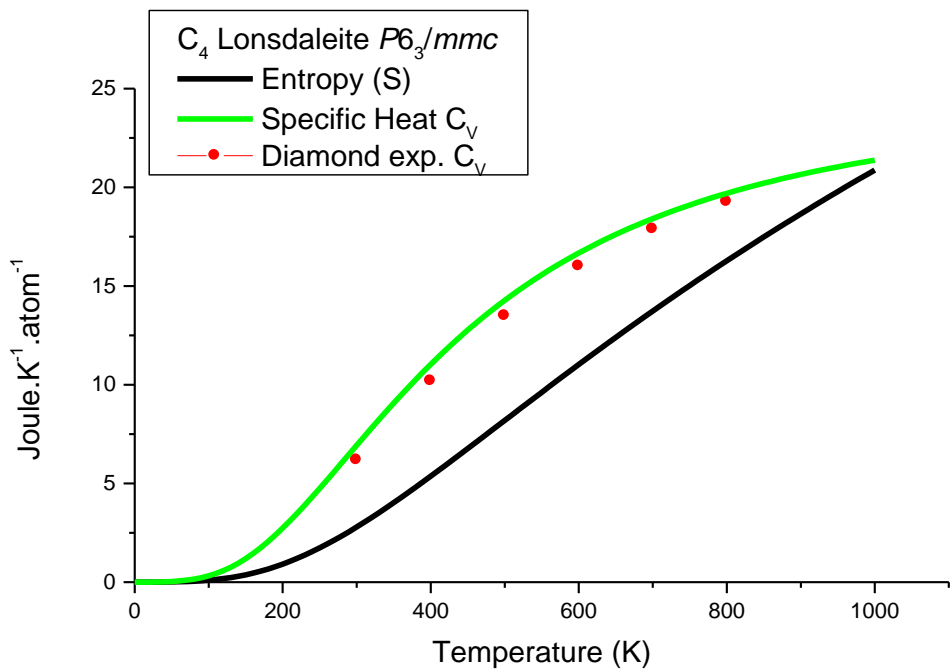
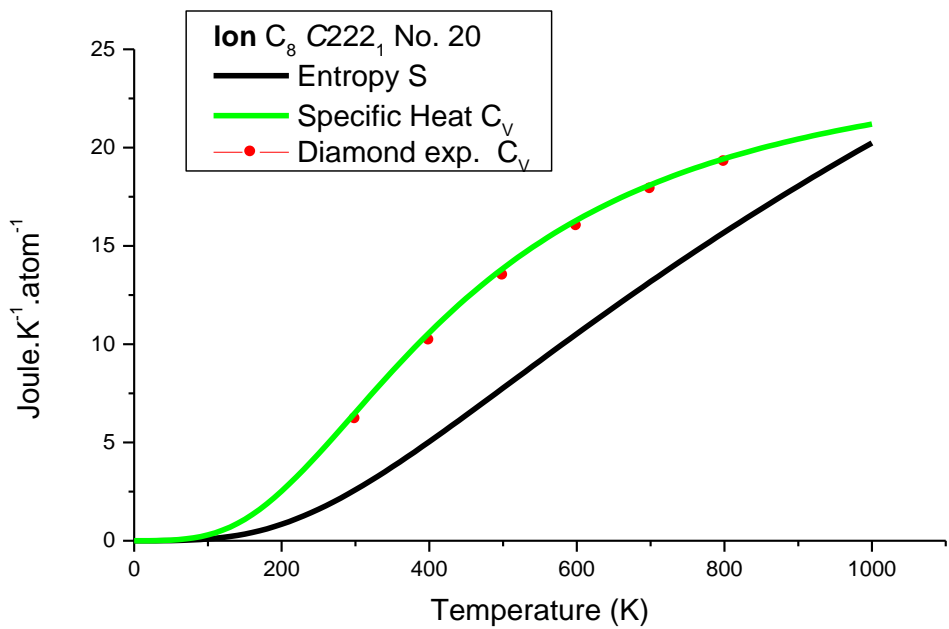

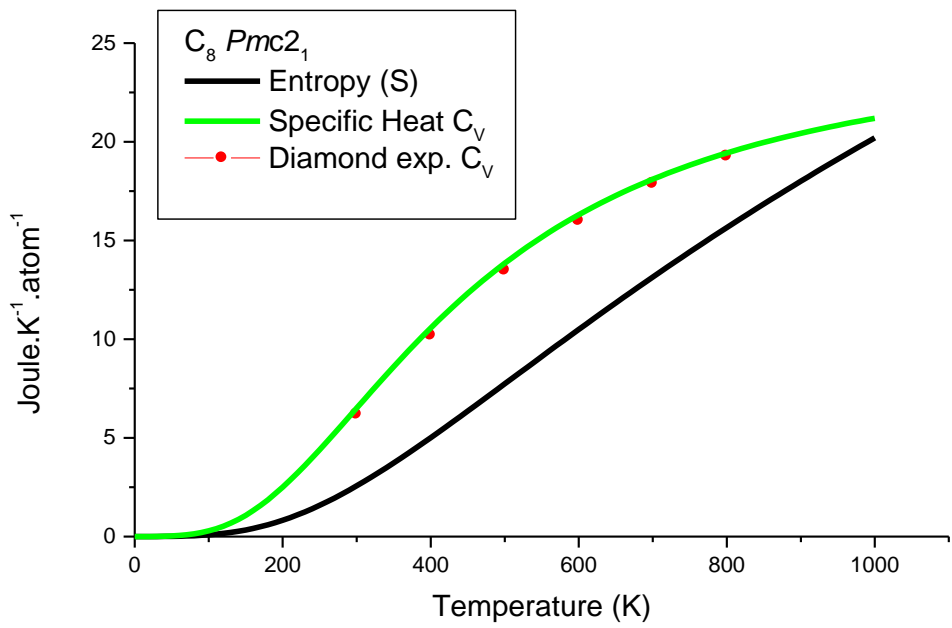

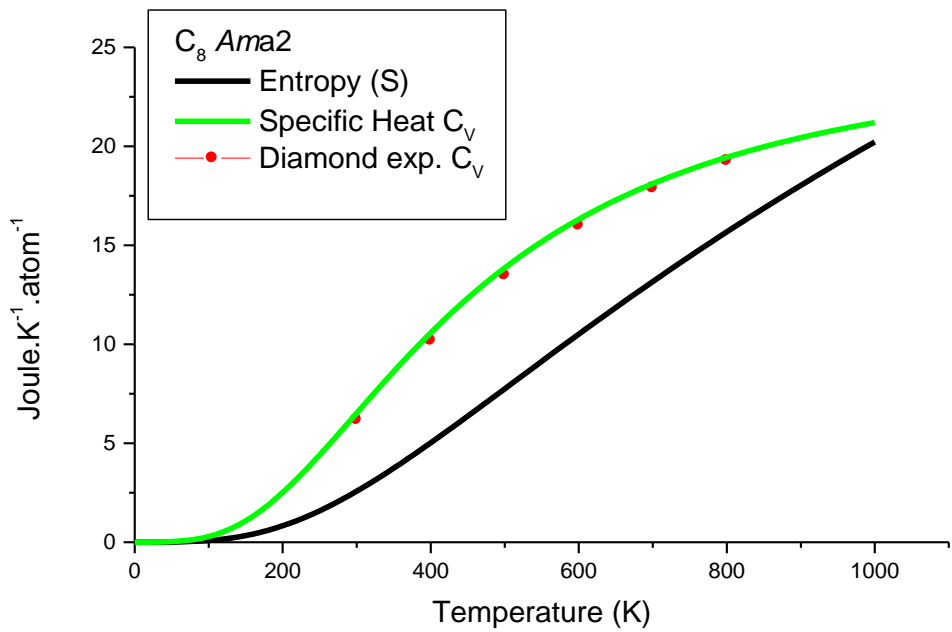

Figure 3. Temperature dependence of the Entropy (S) and the Specific Heat $C_V$ of $C_4$ lonsdaleite and orthorhombic $C_8$ with **lon** topology.

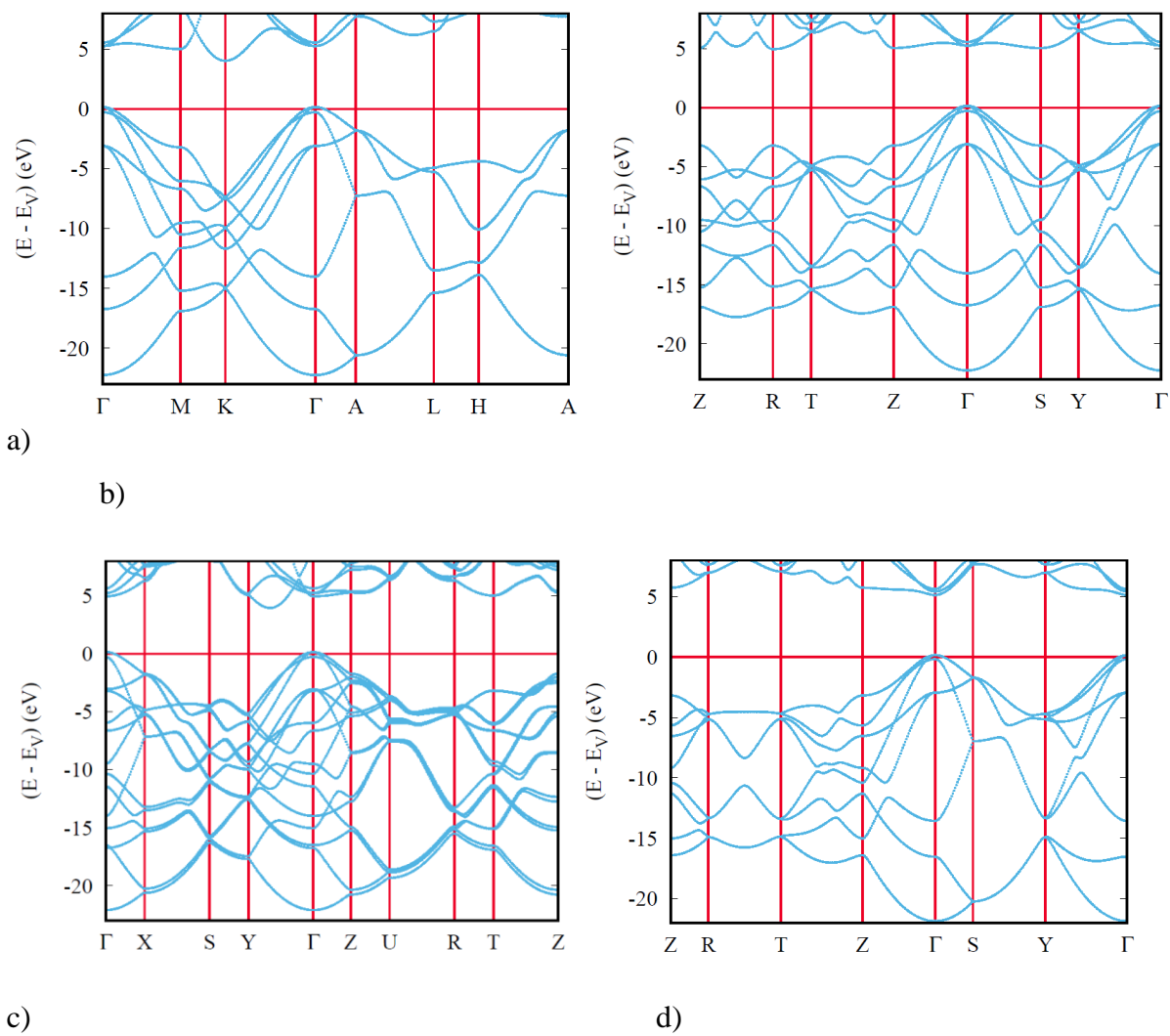

Figure 4. Electronic band structures of $C_4$ Lonsdaleite (a) and orthorhombic $C_8$ with **lon** topology: b) $C_8$ *C*$222_1$, c) $C_8$ *Pmc*$2_1$, and d) $C_8$ *Ama*2.